\documentclass[english,tightenlines,eqsecnum,amsmath,amssymb,aps,prd,superscriptaddress,nofootinbib,showpacs,floats]{article}
\usepackage{epsfig}
\usepackage{babel}
\usepackage{cancel}
\usepackage{authblk}
\usepackage{amsmath,amsfonts,amssymb,multicol}
\usepackage[usenames]{xcolor}
\definecolor{HSH}{rgb}{0.0,0.0,0.9}
\definecolor{HSH1}{rgb}{1,0.0,0.1}
\usepackage[colorlinks,linkcolor=HSH,citecolor=HSH1,bookmarks]{hyperref}
\long\def\/*#1*/{}
\usepackage{color}

{\definecolor{RED}{rgb}{1,0,0}
\definecolor{GREEN}{rgb}{0,1,0}
\definecolor{BLUE}{rgb}{0,0,1}
\usepackage{graphicx,caption,subcaption}

\begin{document}

\title{A connection between Rastall type and $f({\sf R},{\sf T})$ gravities}
\author[1]{Hamid Shabani,\thanks{h.shabani@phys.usb.ac.ir}}
\author[2]{Amir Hadi Ziaie,\thanks{ah.ziaie@maragheh.ac.ir}}

\affil[1]{Physics Department, Faculty of Sciences, University of Sistan and Baluchestan, Zahedan, Iran}
\affil[2]{Research~Institute~for~Astronomy~and~Astrophysics~of~ Maragha~(RIAAM), University of Maragheh,  P.~O.~Box~55136-553,~Maragheh, Iran}
\date{\today}
%
\maketitle
\begin{abstract}
\noindent
Recently a Lagrangian formulation for Rastall Gravity ({\sf RG}) has been proposed in the framework of $f({\sf R},{\sf T})$ gravity~\cite{RasLag}. In the present work we obtain  Lagrangian formulation for the standard Rastall theory assuming the matter content is a perfect fluid with linear Equation of State ({\sf EoS}). We therefore find a relation between the coupling constant of the Lagrangian, the Rastall gravitational coupling constant and the {\sf EoS} parameter. We also propose a Lagrangian for the Generalized Rastall Gravity ({\sf GRG}). In this case the Rastall parameter which is a constant is replaced by a variable one. More exactly, it appears as a function of the Ricci scalar and the trace of the Energy
Momentum Tensor ({\sf EMT}). In both mentioned models, the Lagrangians are constructed by a linear function of {\sf R} and {\sf T}. 
\end{abstract}
\section{Introduction}\label{intro}
In 1972 Rastall has questioned the issue of conservation of {\sf EMT} in the usual form $\nabla_{\mu}{\sf T}^{\mu\nu}=0$~\cite{Rastall1972,Rastall1976}. The basic point in favor of the Rastall proposal is that the conventional expression for {\sf EMT} conservation has only been examined in special relativity and there are no experimental evidences to prove it in non-flat spacetimes. The only non-experimental supporter which validates such an extension is the principle of equivalence~\cite{Rastall1972}. Therefore, Rastall came
to this conclusion that, ``{\it the assumptions from which one derives $\nabla_{\mu}{\sf T}^{\mu\nu}=0$, are
all questionable, so one should not accept this question without further investigation}". He initially proposed
the assumption $\nabla_{\mu}{\sf T}^{\mu\nu}=a^{\nu}$, where the vector field $a_{\nu}$ must vanish in a flat spacetime. He then chose the simplest curvature dependent form for this vector field, i.e., $a_{\nu}=\lambda\nabla_{\nu}{\sf R}$, where ${\sf R}$ is the Ricci Scalar and $\lambda$ is called the Rastall coupling parameter.

Since the advent of Rastall Gravity, various aspects of this theory has been widely explored.
For example, in~\cite{batista2010,capone20101,capone20102,fabris2011,batista2012,batista2013,silva2013,
moradpour20161,moradpour20172}, the cosmological consequences of the theory has been investigated,
G\"{o}del-type solutions have been investigated in~\cite{santos2015} and Brans-Dicke scalar field has been
discussed in~\cite{thiago2014,salako2016}. Static spherically symmetric solutions have been obtained
in~\cite{oliveira2015,oliveira2016,moradpour2016,bronnikov2017,heydarzade2017,lin20191} and authors
of~\cite{majernik2006} have shown that the Rastall proposal is compatible with the Mach's principle. Other
features of Rastall theory can be found in~\cite{yuan2016,moradpour2017,moradpour20177,darabi2018,lobo2018,visser2018,
kumar2018,bamba2018,hansraj2019,moradpour2019,halder2019,ziaie2019,khyllep2019,yu2019}.

Although in the original theory, the Rastall parameter is a constant, some modifications of {\sf RG} has been proposed e.g., in a recent work~\cite{moradpour20171},
the authors have generalized the theory to include a variable Rastall parameter (we hereafter call this theory the Generalized Rastall Gravity {\sf GRG}). They then investigated cosmological consequences of the model $\nabla_{\mu}{\sf T}^{\mu\nu}=\nabla_{\nu}(\lambda'{\sf R})$, where $\lambda^\prime$ is now a function of spacetime coordinates, and predicted a
primary inflationary phase which can exist even in the absence of matter components in a flat FRW spacetime. In a similar work~\cite{das2018}, the authors have concluded that
a cosmological scenario including pre and post inflationary eras can be rendered in the framework of {\sf GRG}. Another formulation for {\sf GRG} has been constructed assuming $a_{\nu}=\lambda\nabla_{\nu}f({\sf R})$~\cite{lin20191} where the authors obtained electrically and magnetically neutral regular black hole solution. Other ideas in this context is to consider possible relations between {\sf GRG} and modified gravity theories such as $f({\sf R},{\sf T})$ and quadratic gravity~\cite{lin20192}. Work along this line has been also carried out in finding other formulations of {\sf RG} whose physical attributes can be investigated in different arenas~\cite{lag,darabi2018}. In this regard, the issue of establishing Lagrangians that provide the field equations of {\sf RG} has become a question of high priority in the literature. Recently, in the context of modified gravity with curvature-matter coupling, a Lagrangian formulation of {\sf RG} has been proposed~\cite{RasLag} and it is shown {\sf RG} can be interpreted as a particular case of these theories of gravity. In the present work we obtain a Lagrangian formulation for the original version of {\sf RG}~\cite{Rastall1972} and {\sf GRG}~\cite{moradpour20171}. In latter case we find that the Rastall parameter is a function of the trace of {\sf EMT}. Also, 
we introduce a new version of {\sf GRG}~\cite{moradpour20171} and obtain the 
corresponding Lagrangian via comparing their 
field equations and also conservation laws to those of $f({\sf R},{\sf T})$ 
gravity~\cite{harko2011}\footnote{$f({\sf R},{\sf T})$ gravity
has been widely studied in the literature, see e.g.,~\cite{shab2013,shab2014,harko2014,zare2016,shab20171,shab20172,shab20173,
shab20174,shab20181,shab20182,singh2018,nagpal2019,sharif2019,moraes2019,ordines2019,
bhatta2019,elizalde2019,baffou2019} for some recent works}. In the case of this new
theory we find that both Rastall gravitational coupling and the
Rastall parameter are functions of the Ricci scalar and the trace
of {\sf EMT}. For both theory we obtain those $f({\sf R},{\sf T})$ Lagrangians that give the field
equations and conservation laws. In the framework of Einstein's gravity, the possibility of dependence of gravitational
coupling to {\sf EMT} trace or Ricci scalar without resorting to least action principle was firstly considered in~\cite{teruel2018} and particularly, cosmological consequences of matter-matter coupling i.e., $\kappa=8\pi G-\alpha{\sf T}$ and matter-curvature coupling i.e., $\kappa=8\pi G+\alpha{\sf R}$ was studied in the context of $\lambda {\sf CDM}$
model. It is however interesting to note that in our Lagrangian formalism for {\sf GRG} we have obtained $\kappa=8\pi G+\partial h({\sf T})/\partial T$,
where $h({\sf T})$ being a function which satisfies some conditions. Incidentally, the form of $h({\sf T})$ is similar to those which have been already considered in a previous
work~\cite{shab20181}. 
\\
The present work is arranged as follows. In Sect.~\ref{sec2} we give a brief review on $f({\sf R},{\sf T})$ gravity and derive the field equations which are required for our discussions. Sec.~\ref{sec3} is devoted to a concise review on the Rastall model and {\sf GRG}. In Sec.~\ref{sec4} we obtain a Lagrangian from $f({\sf R},{\sf T})$ gravity whose metric variation gives the field equation of {\sf RG}. {\sf GRG} and a new version of it 
has been discussed in Sec.~\ref{sec5} and finally, our conclusion is drawn in Sec.~\ref{con}.
\section{Basic equations for $f({\sf R},{\sf T})$ gravity}\label{sec2}
In present section, we review the field equations of $f({\sf R},{\sf T})$ gravity which has been initially
introduced in~\cite{harko2011}. Let us consider the following action
\begin{align}\label{action}
S=\int \sqrt{-{\sf g}} d^{4} x \left[\frac{1}{2\kappa} f\Big{(}{\sf R}, {\sf T}\Big{)}+
{\sf L}^{\rm{(m)}}\right],
\end{align}
where ${\sf G}$ being the gravitational coupling constant and ${\sf T}\equiv {\sf g}^{\mu \nu} {\sf T}_{\mu \nu}$, ${\sf R}$, ${\sf L}^{\rm{(m)}}$ are the trace
of {\sf {\sf EMT}}, the Ricci curvature scalar and the Lagrangian of matter component, respectively. The determinant of the spacetime metric
is denoted by ${\sf g}$, and the units have been set so that $c=1$.
The {\sf EMT} for matter fields reads
\begin{align}\label{Euler-Lagrange}
{\sf T}_{\mu \nu}\equiv-\frac{2}{\sqrt{-{\sf g}}}
\frac{\delta\left[\sqrt{-{\sf g}}{\sf L}^{\rm{(m)}}\right]}{\delta {\sf g}^{\mu \nu}}.
\end{align}
The metric variation of action (\ref{action}) gives the following field equation~\cite{harko2011}
\begin{align}\label{fRTfe1}
&F({\sf R},{\sf T}) {\sf R}_{\mu \nu}-\frac{1}{2} f({\sf R},{\sf T}) {\sf g}_{\mu \nu}+\Big{(} {\sf g}_{\mu \nu}
\square -\triangledown_{\mu} \triangledown_{\nu}\Big{)}F({\sf R},{\sf T})=\nonumber\\
&\Big{(}\kappa-{\mathcal F}({\sf R},{\sf T})\Big{)}{\sf T}_{\mu \nu}-
\mathcal {F}({\sf R},{\sf T})\mathbf
{\Theta_{\mu \nu}},
\end{align}
where
\begin{align}\label{theta1}
\mathbf{\Theta_{\mu \nu}}\equiv {\sf g}^{\alpha \beta}\frac{\delta
{\sf T}_{\alpha \beta}}{\delta {\sf g}^{\mu \nu}}=-2{\sf T}_{\alpha \beta}+
{\sf g}_{\alpha \beta}{\sf L}^{\rm{(m)}}-
2{\sf g}^{\alpha \beta}\frac{\partial^{2}{\sf L}^{\rm{(m)}}}{\partial {\sf g}^{\alpha \beta}\partial {\sf g}^{\mu \nu} },
\end{align}
and we have used the following definitions for the sake of brevity
\begin{align}\label{functiondef1}
{\mathcal F}({\sf R},{\sf T}) \equiv \frac{\partial f({\sf R},{\sf T})}{\partial {\sf T}}~~~~~
~~~~~\mbox{and}~~~~~~~~~~
F({\sf R},{\sf T}) \equiv \frac{\partial f({\sf R},{\sf T})}{\partial {\sf R}}.
\end{align}
The field equation (\ref{fRTfe1}) is applicable when the matter Lagrangian is introduced. Considering then ${\sf L}^{\rm{(m)}}=p$ for a perfect fluid in expression (\ref{theta1}) along with choosing the signature of metric as $(-,+,+,+)$, we obtain
\begin{align}\label{theta2}
\mathbf{\Theta_{\mu \nu}}=-2{\sf T}_{\alpha \beta}+p {\sf g}_{\alpha \beta},
\end{align}
where $p$ is the pressure of fluid. Substituting (\ref{theta2}) into (\ref{fRTfe1}) we get
\begin{align}\label{fRTfe2}
&F({\sf R},{\sf T}) {\sf R}_{\mu \nu}-\frac{1}{2} f({\sf R},{\sf T}) {\sf g}_{\mu \nu}+\Big{(} {\sf g}_{\mu \nu}
\square -\triangledown_{\mu} \triangledown_{\nu}\Big{)}F({\sf R},{\sf T})=\nonumber\\
&\Big{(}\kappa+{\mathcal F}({\sf R},{\sf T})\Big{)}{\sf T}_{\mu \nu}-
\mathcal {F}({\sf R},{\sf T})p {\sf g}_{\mu \nu}.
\end{align}
One can obtain the following covariant equation by applying the Bianchi identity to the field
equation (\ref{fRTfe2}),
\begin{align}\label{relation}
(\kappa +\mathcal {F})\nabla^{\mu}{\sf T}_{\mu \nu}+\frac{1}{2}\mathcal {F}\nabla_{\mu}{\sf T}
+{\sf T}_{\mu \nu}\nabla^{\mu}\mathcal {F}-\nabla_{\nu}(p\mathcal{F})=0,
\end{align}
where we dropped the argument of $\mathcal {F}({\sf R},{\sf T})$ for abbreviation. It is suitable to
rewrite equation (\ref{relation}) in a convenient form for latter purposes, i.e., as
\begin{align}\label{fRTcons}
\nabla^{\mu}{\sf T}_{\mu \nu}=\frac{-\frac{1}{2}\mathcal {F}\nabla_{\mu}{\sf T}
-{\sf T}_{\mu \nu}\nabla^{\mu}\mathcal {F}+\nabla_{\nu}(p\mathcal{F})}{\kappa +\mathcal {F}}.
\end{align}
As it is obvious, in $f({\sf R},{\sf T})$ gravity, {\sf EMT} is not
conserved automatically and this leads to an irreversible particle
creation in cosmology~\cite{harko2014}. In~\cite{harko2014} it is
discussed that such a particle creation is the result of energy
flow from the gravitational field to the matter. Note that, the
only class of solutions for which {\sf EMT} is conserved is of the
form $f({\sf R},{\sf T})={\sf R} +C {\sf T}^{1/2}$ for arbitrary
constant $C$\footnote{For a comprehensive discussion on the
cosmological consequences of these type of models see e.g.,~\cite{shab2013,shab2014}.}.
\section{A brief review on Rastall type gravities}\label{sec3}

In this section we briefly review various versions of Rastall Gravity. {\sf RG} is a relatively simple theory in which both the Rastall parameter and the Rastall gravitational parameter being constants. However, one can also investigate other generalizations ({\sf GRG}) which include one or both mentioned parameter(s) as variable(s). In this sense, in this paper we work on the two cases. The main motivation for such generalizations is to seek for theories with better consistency with the astronomical data. For example, by recent cosmological data one has been informed that there are two accelationary phase in the history of evolution of the universe one of which is absent in the theory of {\sf $\Lambda$CDM}. Along this line, the authors in~\cite{moradpour20171,das2018} have shown that {\sf GRG} leads to inflationary eras in the early and late times of evolution of the universe.

The original version of this theory has been introduced in~\cite{Rastall1972} via modification of the {\sf EMT} conservation as follows
\begin{align}\label{Rascons1}
\nabla_{\mu}{\sf T}^{\mu\nu}=\lambda\nabla^{\nu}{\sf R},
\end{align}
where $\lambda$ is an arbitrary constant. This modification leads to the following field equation
\begin{align}\label{Rasfield1}
{\sf G}_{\mu\nu}+\lambda\kappa'{\sf R}g_{\mu\nu}=\kappa'{\sf T}_{\mu\nu},
\end{align}
where $\kappa'$ is the Rastall gravitational constant. Here, because of different gravitational rule we
have different gravitational coupling. Taking the trace of (\ref{Rasfield1}) and substituting result into field equation (\ref{Rasfield1}) together with using the conservation
rule (\ref{Rascons1}) leads to
\begin{align}
&{\sf G}_{\mu\nu}=\kappa'{\sf T}_{\mu\nu}-
\frac{\lambda\kappa'^{2}}{4\kappa'\lambda-1}{\sf T}{\sf g}_{\mu\nu},\label{Rasfield2}\\
&\nabla_{\mu}{\sf T}^{\mu\nu}=\frac{\lambda\kappa'}{4\kappa'\lambda-1}\nabla^{\nu}{\sf T}.\label{Rascons2}
\end{align}
Equations (\ref{Rasfield2}) and (\ref{Rascons2}) have been written in a suitable form which we need to discuss
them in the next section.
\par
A modified version of Rastall theory i.e., {\sf GRG} has been proposed through using a coordinate dependent Rastall parameter~\cite{moradpour20171}, as
\begin{align}
&\nabla_{\mu}{\sf T}^{\mu\nu}=\nabla^{\nu}(\lambda'{\sf R})\label{Rascons3},\\
&{\sf G}_{\mu\nu}+\lambda'\kappa''{\sf R}g_{\mu\nu}=\kappa''{\sf T}_{\mu\nu},\label{Rasfield3}
\end{align}
where we have used a different gravitational coupling. The trace of field equation (\ref{Rasfield3}) leaves us with
a relation between three variables ${\sf T}$, {\sf R} and $\lambda'$, as
\begin{align}\label{trace2}
(4\kappa''\lambda'-1){\sf R}=\kappa''{\sf T},
\end{align}
with the help of which equation (\ref{Rascons3}) gives
\begin{align}
&\nabla_{\mu}{\sf T}^{\mu\nu}=
\nabla^{\nu}\left(\frac{\kappa''\lambda'{\sf T}}{4\kappa''\lambda'-1}\right)\label{Rascons4}.
\end{align}
\section{A suggestion for Lagrangian of Rastall gravities from $f({\sf R},{\sf T})$ theory}\label{sec4}
In this section we try to find an acceptable Lagrangian from $f({\sf R},{\sf T})$ gravity for 
Rastall type gravities. We begin with the original Rastall model~\cite{Rastall1972}. Since in the 
Rastall field equation (\ref{Rasfield2}) there appear no derivatives of the Ricci scalar, a comparison to 
equation (\ref{fRTfe2}) suggests a function of type 
\begin{align}\label{e0}
f({\sf R},{\sf T})={\sf R}+\alpha{\sf T},
\end{align}
where $\alpha$ being some coupling constant. Substituting this function into (\ref{fRTfe2}) gives
\begin{align}\label{e1}
{\sf G}_{\mu\nu}=(\kappa+\alpha){\sf T}_{\mu\nu}+\alpha\left(\frac{{\sf T}}{2}-p\right){\sf g}_{\mu\nu}.
\end{align}
Using the perfect fluid {\sf EoS}, $p=w\rho$, leads to the relation ${\sf T}=-\rho+3p=(3w-1)\rho$ from which we have
\begin{align}\label{e2}
{\sf G}_{\mu\nu}=(\kappa+\alpha){\sf T}_{\mu\nu}+
\frac{\alpha(w-1)}{2 (3 w-1)}{\sf T}{\sf g}_{\mu\nu}.
\end{align}
Therefore, equation (\ref{e2}) gives (\ref{Rasfield2}) provided that the following equalities hold
\begin{align}\label{e3}
\kappa'=\kappa+\alpha~~~~~~~\mbox{and}~~~~~~
-\frac{\lambda\kappa'^{2}}{4\kappa'\lambda-1}=\frac{\alpha(w-1)}{2(3w-1)},
\end{align}
from which we get the coupling constant in Lagrangian (\ref{e0}) as
\begin{align}\label{e4}
\alpha=\frac{2(1-3w)\lambda\kappa'^{2}}{(w-1)(4\kappa'\lambda-1)}.
\end{align}
It is easy to check that for solution (\ref{e4}) and function (\ref{e0}),
two equations (\ref{fRTcons}) and (\ref{Rascons2}) become identical. 
In this case for choice (\ref{e0})
we obtain
\begin{align}\label{e5}
\nabla_{\mu}{\sf T}^{\mu \nu}=
\frac{\alpha}{2(\kappa+\alpha)}\frac{w-1}{1-3w}\nabla^{\nu}{\sf T}=
\frac{\lambda\kappa'}{4\kappa'\lambda-1}\nabla^{\nu}{\sf T}.
\end{align}
Therefore, introducing a Lagrangian formulation from $f({\sf R},{\sf T})$ requires a change in the Rastall
gravitational constant, $\kappa'$. As a result, we can say {\sf RG} can be understood from
 $f({\sf R},{\sf T})$ Lagrangian provided that results (\ref{e3}) are satisfied and also function
(\ref{e0}) is utilized.
\section{Generalization of Rastall Gravity and a proposal for it's Lagrangian}\label{sec5}
In this section we obtain a Lagrangian formulation for {\sf GRG} and introduce a new version of 
{\sf GRG} and propose its corresponding Lagrangian. We first suppose that, in general, the gravitational coupling $\kappa''$ 
in equation (\ref{Rasfield3}) is a variable. This assumption gives a new version of {\sf GRG} for which the Bianchi 
identity cannot be satisfied, obviously. In the case of a varying $\kappa''$ applying 
the Bianchi identity to equation (\ref{Rasfield3}) gives
\begin{align}\label{rela1}
\kappa''\nabla_{\mu}(\lambda' {\sf R})+\lambda' {\sf R}\nabla_{\mu}\kappa''=
\kappa''\nabla^{\nu}{\sf T}_{\mu\nu}+{\sf T}_{\mu\nu}\nabla^{\nu}\kappa''.
\end{align}
As it can be seen, to reach the original equation (\ref{Rascons3}), there is a constraint which must hold,
i.e.,
\begin{align}\label{cons1}
\lambda' {\sf R}\nabla_{\mu}\kappa''-{\sf T}_{\mu\nu}\nabla^{\nu}\kappa''=0.
\end{align}
Now, substituting the Lagrangian 
\begin{align}\label{l2}
f({\sf R},{\sf T})={\sf R}+h({\sf T}),
\end{align}
into the field equation (\ref{fRTfe2}) we find
\begin{align}\label{e7}
{\sf G}_{\mu\nu}=\left(\kappa+\frac{d h}{d {\sf T}}\right){\sf T}_{\mu\nu}+
\left(\frac{h}{2}-p\frac{d h}{d {\sf T}}\right){\sf g}_{\mu\nu}.
\end{align}
In this case, equality of two equations (\ref{Rasfield3}) (for a varying $\kappa''$) and (\ref{e7}) requires that the following relations be satisfied
\begin{align}\label{e8}
&\kappa''=\kappa+\frac{d h}{d {\sf T}},\nonumber\\
&\lambda'=-\frac{\frac{h}{2}-p\frac{d h}{d {\sf T}}}{{\sf R}\left(\kappa+\frac{d h}{d {\sf T}}\right)}=\frac{\frac{h}{2}-p \frac{d h}{d {\sf T}}}{\left(\frac{d h}{d {\sf T}}+\kappa \right)
\left[{\sf T} \left(\frac{d h}{d {\sf T}}+\kappa \right)+4 \left(\frac{h}{2}-p \frac{d h}{d {\sf T}}\right)\right]}.
\end{align}
Next, we note that the constraint (\ref{cons1}) should be considered for relations (\ref{e8}). Substituting then for $\lambda'$ from (\ref{e8}) into (\ref{cons1}) gives
\begin{align}\label{e9}
\left(\frac{h}{2}-\frac{d h}{d {\sf T}}p\right)\nabla_{\mu}\kappa''
+\Big(\kappa+\frac{d h}{d {\sf T}}\Big){\sf T}_{\mu\nu}\nabla^{\nu}\kappa''=0.
\end{align}
Equation (\ref{e9}) is a constraint that must be satisfied. Assuming a homogeneous 
gravitational coupling parameter $\kappa''$ along with a perfect fluid for the matter content, equation (\ref{e9}) reduces to the following equation\footnote{Note that in the case of a homogeneous coupling parameter $\kappa''$, the time derivative of $\kappa''$ would appear in both terms of (\ref{e9}). Also, we have used ${\sf T}^{t}_{t}=-\rho={\sf T}/(1-3w)$ and $p=w\rho=w{\sf T}/(3w-1)$ to obtain equation (\ref{e10}).}
\begin{align}\label{e10}
\frac{d h}{d {\sf T}}{\sf T}-\frac{3w-1}{2(w+1)}h+\frac{\kappa}{w+1}{\sf T}=0,
\end{align}
for which the solution reads
\begin{align}\label{e100}
h({\sf T})=\frac{2 \kappa}{w-3} {\sf T}+\beta{\sf T}^{\frac{3 w-1}{2 (w+1)}},
\end{align}
where $\beta$ is an integration constant\footnote{One can find some interesting features of functions
of this form in~\cite{shab20181,shab20182}.}. Note that in the case of constraint (\ref{e9}) one gets 
$\kappa''\lambda'=1/3(1+w)$. Therefore, it suffices to use function (\ref{e100}) as well as
gravitational coupling $\kappa''$ in (\ref{e8}) to obtain {\sf GRG} equations (\ref{Rascons3}) and
(\ref{Rasfield3}) from (\ref{l2}). However, in our approach, both the Rastall parameter and gravitational 
coupling have to be variables instead of constant ones, contrary to the case of {\sf GRG} which originally 
introduced in~\cite{moradpour20171}. 

It remains to check whether the equality of equations (\ref{fRTcons}) and (\ref{Rascons3}) leads to 
solution (\ref{e100}). In the case of Lagrangian (\ref{l2}), we can rewrite equation (\ref{fRTcons}) as
\begin{align}\label{fRTcons3}
\nabla^{\mu}{\sf T}_{\mu \nu}=\frac{1}{(3w-1)(\kappa+\mathcal {F})}\left[\frac{1-w}{2}
\mathcal {F}+(1+w){\sf T}\frac{d \mathcal {F}}{d {\sf T}}\right]\nabla_{\nu}{\sf T},
\end{align}
while for equation (\ref{Rascons4}) we obtain
\begin{align}\label{Rascons5}
\nabla^{\mu}{\sf T}_{\mu \nu}=
\left(\frac{x}{4x-1}-\frac{{\sf T}}{(4x-1)^2}\frac{d x}{d {\sf T}}\right)\nabla_{\nu}{\sf T},
\end{align}
where, $x\equiv\kappa''\lambda'$. One can easily check that 
equations (\ref{fRTcons3}) and (\ref{Rascons5}) lead to the same result provided that 
equation (\ref{e10}) holds. More precisely, in this case we have
\begin{align}\label{e30}
\frac{d^2 h}{d {\sf T}^2}{\sf T}-\frac{w-3}{2(w+1)}\frac{d h}{d {\sf T}}{\sf T}+
\frac{\kappa}{w+1}=0,
\end{align}
which is the derivative of equation (\ref{e10}) with respect to the trace of {\sf EMT}.

Now, we consider equations (\ref{cons1}) and (\ref{e8}) for a constant gravitational coupling 
and again utilize the function (\ref{l2}). Clearly, (\ref{cons1}) is automatically satisfied. The upper relation 
in (\ref{e8}) gives the solution $h({\sf T})=(\kappa''-\kappa){\sf T}+\Lambda$, for arbitrary constant 
$\Lambda$. Substituting this solution into the lower relation in (\ref{e8}) yields
\begin{align}\label{e11}
\lambda'=\frac{1}{4} \left(\frac{(3 w-1){\sf T}}{\left[2(w-1)\kappa +(3-5 w)\kappa''\right]{\sf T}+2 \Lambda(1-3w)}+\frac{1}{\kappa''}\right).
\end{align}
Next, we must ensure that both conservation equations (\ref{fRTcons}) and (\ref{Rascons3}) provide the same
result. Equation (\ref{fRTcons}) for Lagrangian $h({\sf T})=(\kappa''-\kappa){\sf T}+\Lambda$ along with perfect fluid assumption leads to
\begin{align}\label{fRTcons2}
\nabla^{\mu}{\sf T}_{\mu \nu}=\frac{1-w}{2(3w-1)}\frac{\mathcal {F}}{\kappa+\mathcal {F}}\nabla_{\nu}{\sf T},
\end{align}
while (\ref{Rascons3}) gives
\begin{align}\label{Rastallcons2}
\nabla^{\mu}{\sf T}_{\mu \nu}=\kappa''\left(\frac{\lambda'}{4\kappa''\lambda'-1}-\frac{Td\lambda'/d{\sf T}}{(4\kappa''\lambda'-1)^2}\right)\nabla_{\nu}{\sf T}.
\end{align}
Equality of equations (\ref{fRTcons2}) and (\ref{Rastallcons2}) leads to a differential equation for which the solution
(\ref{e11}) is recovered provided that the free parameter within the solution of this differential equation is chosen as $-\Lambda/2\kappa''^2$.

Therefore, {\sf GRG} is equivalent to $f({\sf R},{\sf T})={\sf R}+(\kappa''-\kappa){\sf T}+\Lambda$ gravity. In this case the Rastall parameter
is a function of the trace of {\sf EMT} and depends on the Rastall gravitational constant $\kappa''$.
\section{Concluding remarks}\label{con}
In this paper we considered a possible connection between {\sf RG} and its generalizations, and Lagrangians 
of $f({\sf R}, {\sf T})$ gravity theories. We first obtained a relation between Rastall field equations 
and the linear Lagrangian $f({\sf R}, {\sf T})={\sf R}+\alpha{\sf T}$. In this case one obtains the 
Rastall field equations as well as the Rastall {\sf EMT} conservation rule from the mentioned Lagrangian.

In the next step, we investigated the Lagrangian of a version of {\sf GRG} which has been introduced 
in~\cite{moradpour20171}. It is found that for this version of {\sf GRG} to be derived from
 $f({\sf R}, {\sf T})$ gravity, the Rastall parameter could be running. For this case, we chose the
 Lagrangian $f({\sf R},{\sf T})={\sf R}+h({\sf T})$ and compared the field equations derived from this 
Lagrangian to those of the new version of {\sf GRG}. We then found the Rastall parameter and the Rastall gravitational 
coupling parameter in terms of some functions of the Ricci scalar and the trace of {\sf EMT}. Applying 
the Bianchi identity on the related field equation of this version of {\sf GRG} leads to covariant 
constraint (\ref{e9}) which must be satisfied. Hence, solving this constraint gives the 
function (\ref{e100}). Also, it was verified that the {\sf EMT} conservation equation (\ref{fRTcons}) leads
to equation (\ref{Rascons3}) for solution (\ref{e100}).

Finally, we have shown that {\sf GRG} is equivalent to 
$f({\sf R},{\sf T})={\sf R}+(\kappa''-\kappa){\sf T}+\Lambda$ gravity with a Rastall parameter which 
is a function of the trace of {\sf EMT} and a constant Rastall gravitational coupling. It is notable that, 
recently, similar considerations has been performed for only {\sf RG}~\cite{RasLag}. However, in the 
present work we also have considered {\sf GRG} as well as it's {\sf EMT} conservation rule and an 
extension of it in a more straightforward way.


\begin{thebibliography}{99}
\bibitem{RasLag} W. A. G. De Moraes and A. F. Santos, Gen. Relativ. Grav. {\bf 51}, 167 (2019).
\bibitem{Rastall1972} P. Rastall, Phys. Rev. D {\bf 6}, 3357 (1972).
\bibitem{Rastall1976} P. Rastall, Can. J. Phys. {\bf54}, 66 (1976).
\bibitem{batista2010} C. E. M. Batista, J.C. Fabris and M. Hamani Daouda, Nuovo Cim. B {\bf125}, 957 (2010).
\bibitem{capone20101} M. Capone, V. F. Cardone and M. L. Ruggiero, Nuovo Cim. B {\bf125}, 1133 (2010).
\bibitem{capone20102} M Capone, V. F. Cardone and M. L. Ruggiero, J. Phys. Conf. Ser. {\bf222}, 012012 (2010).
\bibitem{fabris2011} J. C. Fabris, T. C. C. Guio, M. Hamani Daouda and O. F. Piattella, Grav. Cosmol. {\bf17}, 259 (2011).
\bibitem{batista2012} C. E. M. Batista, M. H. Daouda, J. C. Fabris, O. F. Piattella and D. C. Rodrigues, Phys. Rev. D {\bf 85}, 084008  (2012).
\bibitem{batista2013} C. E. M. Batista, J. C. Fabris, O. F. Piattella, A. M. Velasquez-Toribio, Eur. Phys. J. C {\bf73}, 2425 (2013).
\bibitem{silva2013} G. F. Silva, O. F. Piattella, J. C. Fabris, L. Casarini and T. O. Barbosa, Grav. Cosmol. {\bf 19}, 156 (2013).
\bibitem{moradpour20161} H. Moradpour, Phys. Lett. B {\bf757}, 187 (2016).
\bibitem{moradpour20172} H. Moradpour, A. Bonilla, E. M. C. Abreu, and J. A. Neto, Phys. Rev. D {\bf96}, 123504 (2017).
\bibitem{santos2015} A. F. Santos and S. C. Ulhoa, Mod. Phys. Lett. A {\bf30}, 1550039 (2015).
\bibitem{thiago2014}  T. R. P. Caramês et. al., EPJC {\bf74}, 3145 (2014).
\bibitem{salako2016} I. G. Salako, M. J. S. Houndjo and A. Jawad, Int. J. Mod. Phys. D {\bf25}, 1650076  (2016).
\bibitem{oliveira2015} A. M. Oliveira, H. E. S. Velten, J. C. Fabris and L. Casarini, Phys. Rev. D {\bf92}, 044020  (2015).
\bibitem{oliveira2016} A. M. Oliveira, H. E. S. Velten and J. C. Fabris, Phys. Rev. D {\bf93}, 124020 (2016).
\bibitem{moradpour2016} H. Moradpour and I. G. Salako, Adv. High Energy Phys. {\bf2016}, 3492796 (2016).
\bibitem{bronnikov2017} K. A. Bronnikov, J. C. Fabris, O. F. Piattella, E. C. Santos, Gen. Rel. Grav. {\bf48}, 162 (2016).
\bibitem{heydarzade2017} Y. Heydarzade, H. Moradpour and F. Darabi, Can. J. Phys. {\bf95}, 1253 (2017).
\bibitem{lin20191} K. Lin and W.-L. Qian, Chi. Phys. C {\bf43}, 083106 (2019).
\bibitem{majernik2006}  V. Majernik and L. Richterek, arxiv: gr-qc/0610070.
\bibitem{yuan2016} F.-F. Yuan and P. Huang, Class. Quant. Grav. {\bf34}, 077001 (2017).
\bibitem{moradpour2017} H. Moradpour, C. Corda, I. Licata, arxiv: gen-ph/1711.01915.
\bibitem{moradpour20177} H. Moradpour, N. Sadeghnezhad and S. H. Hendi, Can. J. Phys, {\bf95}, 1257 (2017).
\bibitem{darabi2018} F. Darabi, H. Moradpour, I. Licata, Y. Heydarzade and C. Corda, Eur. Phys. J. C {\bf78}, 25 (2018).
\bibitem{lobo2018} I. P. Lobo, H. Moradpour, J. P. Morais Graça and I. G. Salako,  Int. J. Mod. Phys. D {\bf27}, 1850069 (2018).
\bibitem{visser2018} M. Visser,  Phys. Let. B {\bf782}, 83 (2018).
\bibitem{kumar2018} R. Kumar and S. G. Ghosh, Eur. Phys. J. C {\bf78}, 750 (2018).
\bibitem{bamba2018} K. Bamba, A. Jawad, S. Rafique and Hooman Moradpour, Eur. Phys. J. C {\bf78}, 986 (2018).
\bibitem{hansraj2019} S. Hansraj, A. Banerjee and P. Channuie, Ann. Phys. {\bf400}, 320 (2019).
\bibitem{moradpour2019} H. Moradpour and M. Valipour, arxiv: gen-ph/1901.05288.
\bibitem{halder2019} S. Halder, S. Bhattacharya, S. Chakraborty, Mod. Phys. Lett. A {\bf34}, 1950095 (2019).
\bibitem{ziaie2019} A. H. Ziaie, H. Moradpour, S. Ghaffari,  Phys. Let. B {\bf793}, 276 (2019).
\bibitem{khyllep2019} W. Khyllep and J. Dutta, Phys. Let. B {\bf797}, 134796 (2019).
\bibitem{yu2019} Z.-X. Yu and Hao Wei, arXiv: gr-qc/1907.12517.
\bibitem{moradpour20171} H. Moradpour, Y. Heydarzade, F. Darabi and Ines G. Salako, Eur. Phys. J. C {\bf77}, 259 (2017).
\bibitem{das2018} D. Das, S. Dutta and S. Chakraborty, Eur. Phys. J. C {\bf78}, 810 (2018).
\bibitem{lin20192} K. Lin, Y. Liu and W.-L. Qian, Gen. Rel. Grav. {\bf51}, 62 (2019).
\bibitem{lag} L. L. Smalley, Il Nuovo Cimento B 80, 42 (1984);\\ V. Dzhunushaliev and H. Quevedo, Gravit. Cosmol. {\bf 23}, 280 (2017);\\ I. Licata, H. Moradpour and C. Corda, Int. J. Geom. Methods Mod. Phys. 14, 1730003
(2017);\\ R. V. d-Santos and J. A. C. Nogales, arXiv:1701.08203;\\
H. Moradpour, I. Licata, C. Corda, I. G. Salako, Mod. Phys. Lett A
34 (13), 1950096 (2019).
\bibitem{harko2011} T. Harko, F. S. N. Lobo, S. Nojiri and , S. D. Odintsov, Phys. Rev. D {\bf 84}, 024020 (2011).
\bibitem{shab2013} H. Shabani and M. Farhoudi, Phys. Rev. D {\bf 88}, 044048 (2013).
\bibitem{shab2014} H. Shabani and M. Farhoudi, Phys. Rev. D {\bf 90}, 044031 (2014).
\bibitem{harko2014}  T. Harko, Phys. Rev. D {\bf 90}, 044048 (2013).
\bibitem{zare2016} R. Zaregonbadi, M. Farhoudi and N. Riazi, Phys. Rev. D {\bf 94}, 084052 (2016)
\bibitem{shab20171} H. Shabani and A. H. Ziaie, Eur. Phys. J.  C {\bf 77}, 31 (2017).
\bibitem{shab20172} H. Shabani and A. H. Ziaie, Eur. Phys. J. C {\bf 77}, 282 (2017).
\bibitem{shab20173} H. Shabani, Int. J. Mod. Phys. D. {\bf 26}, 1750120 (2017).
\bibitem{shab20174}  H. Shabani and A. H. Ziaie,  Eur. Phys. J. C {\bf 77}, 507 (2017).
\bibitem{shab20181} H. Shabani and A. H. Ziaie, Int. J. Mod. Phys. A {\bf33}, 1850050 (2018).
\bibitem{shab20182}  H. Shabani and A. H. Ziaie,  Eur. Phys. J. C {\bf 78}, 397 (2018).
\bibitem{singh2018} J. K. Singh, K. Bamba, R. Nagpal and S. K. J. Pacif, Phys. Rev. D {\bf 99}, 1233536 (2018).
\bibitem{nagpal2019} R. Nagpal, J. K. Singh, A. Beesham and H. Shabani, Ann. Phys. {\bf 405}, 234 (2019).
\bibitem{sharif2019} M. Sharif and A. Siddiqa, Gen. Rel. Grav. {\bf 51}, 74 (2019).
\bibitem{moraes2019} P. H. R. S. Moraes, Eur. Phys. J. C {\bf 79},674 (2019).
\bibitem{ordines2019} T. M. Ordines and E. D. Carlson, Phys. Rev. D {\bf 99}, 104059 (2019).
\bibitem{bhatta2019} P. K. Sahoo and S. Bhattacharjee, arxiv: 1907.13460.
\bibitem{elizalde2019} E. Elizalde and M. Khurshudyan, arxiv: gr-qc/1909.11037.
\bibitem{baffou2019} E. H. Baffou, M. J. S. Houndjo, D. A. Kanfon and I. G. Salako, Eur. Phys. J. C {\bf 79}, 112 (2019).
\bibitem{teruel2018} G. R. P. Teruel,  Eur. Phys. J. C {\bf 78}, 660 (2018).
\end{thebibliography}
\end{document}